\documentclass[11pt,twoside]{article}

\usepackage{asp2006}
\usepackage{epsf}
\usepackage{psfig}
\usepackage{lscape}

\begin{document}
\title{Magnetic-field fluctuations around the radio galaxy 3C 31}   %%% Fill in title
\author{R.A. Laing}   %%% Fill in author names
\affil{ESO, Karl-Schwarzschild-Stra\ss e 2, 85748 Garching-bei-M\"{u}nchen,
  Germany}    
%%% Fill in author affiliations
\author{A.H. Bridle}   %%% Fill in author names
\affil{NRAO, 520 Edgemont Road, Charlottesville, VA
22903-2475, U.S.A.}    %%% Fill in author affiliations

\begin{abstract} %%% Abstract to run on from here.
We present an analysis of the magnetic-field fluctuations in the magnetoionic
medium in front of the radio galaxy 3C 31 derived from rotation-measure (RM)
fits to high-resolution polarization images.  We first show that the Faraday
rotation must be due primarily to a foreground medium.  We determine the RM
structure functions for different parts of the source and infer that the
simplest form for the power spectrum is a power law with a high-frequency
cutoff. We also present three-dimensional simulations of RM produced by a
tangled magnetic field in the hot plasma surrounding 3C 31, and show that 
the observed RM distribution is consistent with a spherical plasma distribution
in which the radio source has produced a cavity.
\end{abstract}

%%% MAIN BODY OF TEXT GOES HERE. CONSULT "INSTRUCTIONS FOR AUTHORS USING
%%% LATEX2E MARKUP", SECTIONS 2.3-2.6 FOR HELP WITH EQUATIONS, FIGURES,
%%% AND TABLES.

%\section{}   %%% Top level section head (remove "%" symbol)
%\subsection{}   %%% Second level section head (remove "%" symbol)
%\subsubsection{}   %%% Lowest level section head (remove "%" symbol)
%\section*{}    %%% Unnumbered top level section head (remove "%" symbol)
%\subsection*{}   %%% Unnumbered second level section head (remove "%" symbol)

\begin{figure}
\plotone{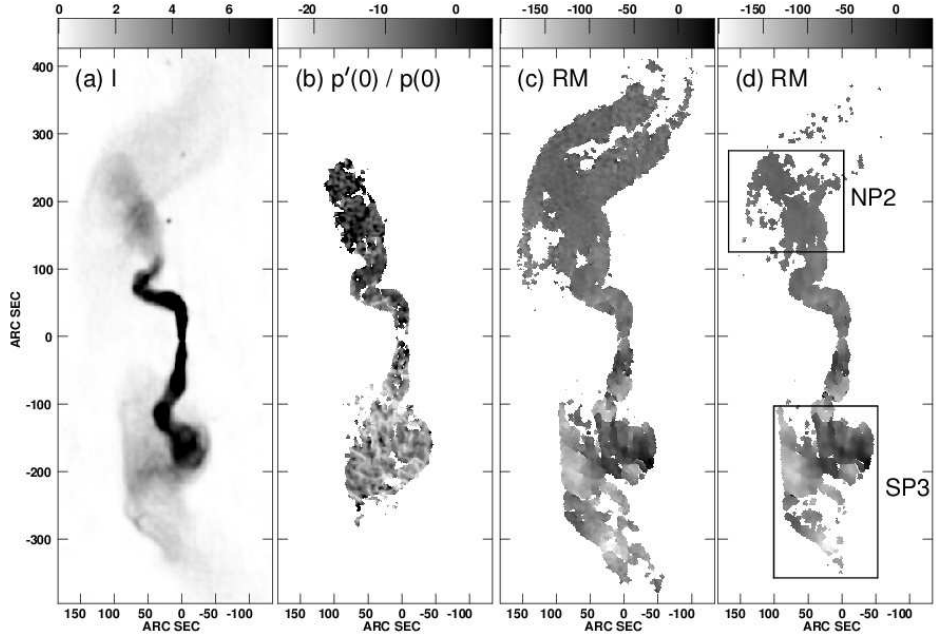}
\caption{Images at 5.5\,arcsec resolution. (a) Mean of four total-intensity
  images at frequencies from 1.4 -- 1.7\,GHz. (b) Normalized polarization
  gradient $p^\prime(0)/p(0)$, in m$^{-2}$. (c) RM from a fit to 4 images at
  frequencies between 1.4 and 1.7 GHz, in rad\,m$^{-2}$. (d) As (c), but for 5
  frequencies from 1.4 -- 5.0 GHz.\label{fig:rmlo}}
\end{figure}

\begin{figure}
\plotfiddle{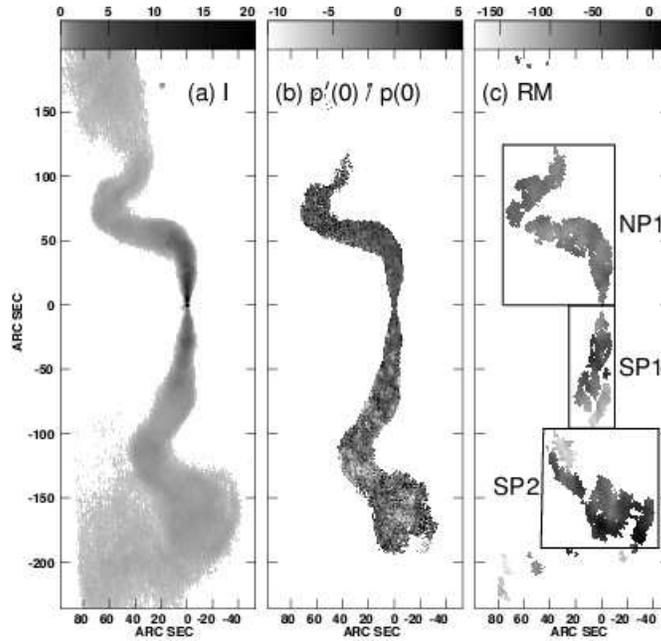}{7.5cm}{0}{45}{45}{-130}{-20 }
\caption{Images at 1.5 arcsec resolution. (a) Mean of 4 total-intensity images
  at frequencies from 1.4 -- 1.7 GHz. (b) Normalized polarization gradient
  $p^\prime(0)/p(0)$, in m$^{-2}$. (c) RM in rad\,m$^{-2}$ from a fit to 6
  frequencies between 1.4 and 8.4 GHz using the {\sc pacerman} algorithm
  \citep{Pacerman}.
  \label{fig:rmhi}}
\end{figure}

\section{Observed Faraday rotation and depolarization}

Our knowledge of the structure and origin of magnetic fields in elliptical
galaxies, groups and clusters is still rudimentary, but Faraday rotation of
linearly-polarized radio emission can be used to probe the fields in ionized
foreground gas.  Here, we present an analysis of the magnetic-field fluctuations
in the magnetoionic medium in front of the FR\,I radio galaxy 3C\,31 (z =
0.0169) derived from rotation-measure (RM) fits to high-resolution polarization
images.

Our analysis is based on VLA observations at 6 frequencies in the range 1.4 --
8.4\,GHz with resolutions of 5.5 and 1.5\,arcsec FWHM (Figs~\ref{fig:rmlo} and
\ref{fig:rmhi}). We show images of normalized polarization gradient
$p^\prime(0)/p(0)$ from fits to $p(\lambda^2) = p(0) + p^\prime(0)\lambda^2$,
where $p(\lambda^2)$ is the degree of polarization at wavelength $\lambda$ and a
prime denotes differentiation with respect to $\lambda^2$, together with RM
images derived from fits to $\chi(\lambda^2) = \chi(0) + {\rm RM}\lambda^2$ at 4
-- 6 wavelengths, where $\chi$ is the {\bf E}-vector position angle.

The residual depolarization at 1.5\,arcsec resolution is very small and the
rotation is accurately proportional to $\lambda^2$, indicating almost
completely resolved foreground rotation. There is a large asymmetry across the
nucleus: the lobe with the brighter jet shows a much smaller RM fluctuation
amplitude than the counter-jet lobe on all scales, qualitatively as expected
from relativistic jet models \citep{L88}.

\section{Structure functions and power spectra}

\begin{figure}
\plotfiddle{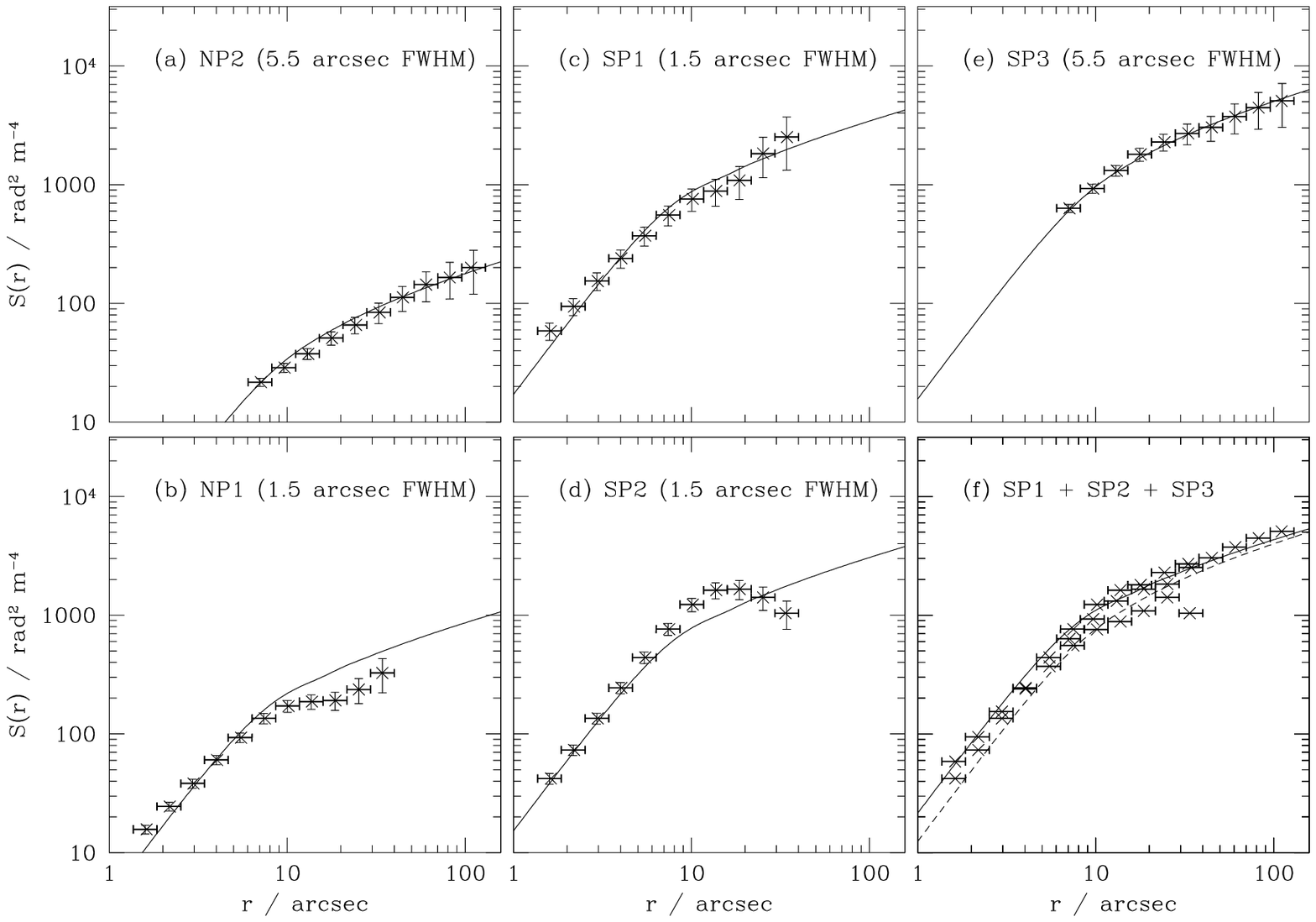}{6.5cm}{0}{60}{60}{-150}{-150}
\caption{(a) -- (f): plots of the RM structure function for the regions
  indicated in Figs~\ref{fig:rmlo} and \ref{fig:rmhi}.  The curves are the
  predictions for the cut-off power law power spectrum described in the text,
  including the effects of the convolving beam.\label{fig:sf}}
\end{figure}

\begin{figure}
\plotone{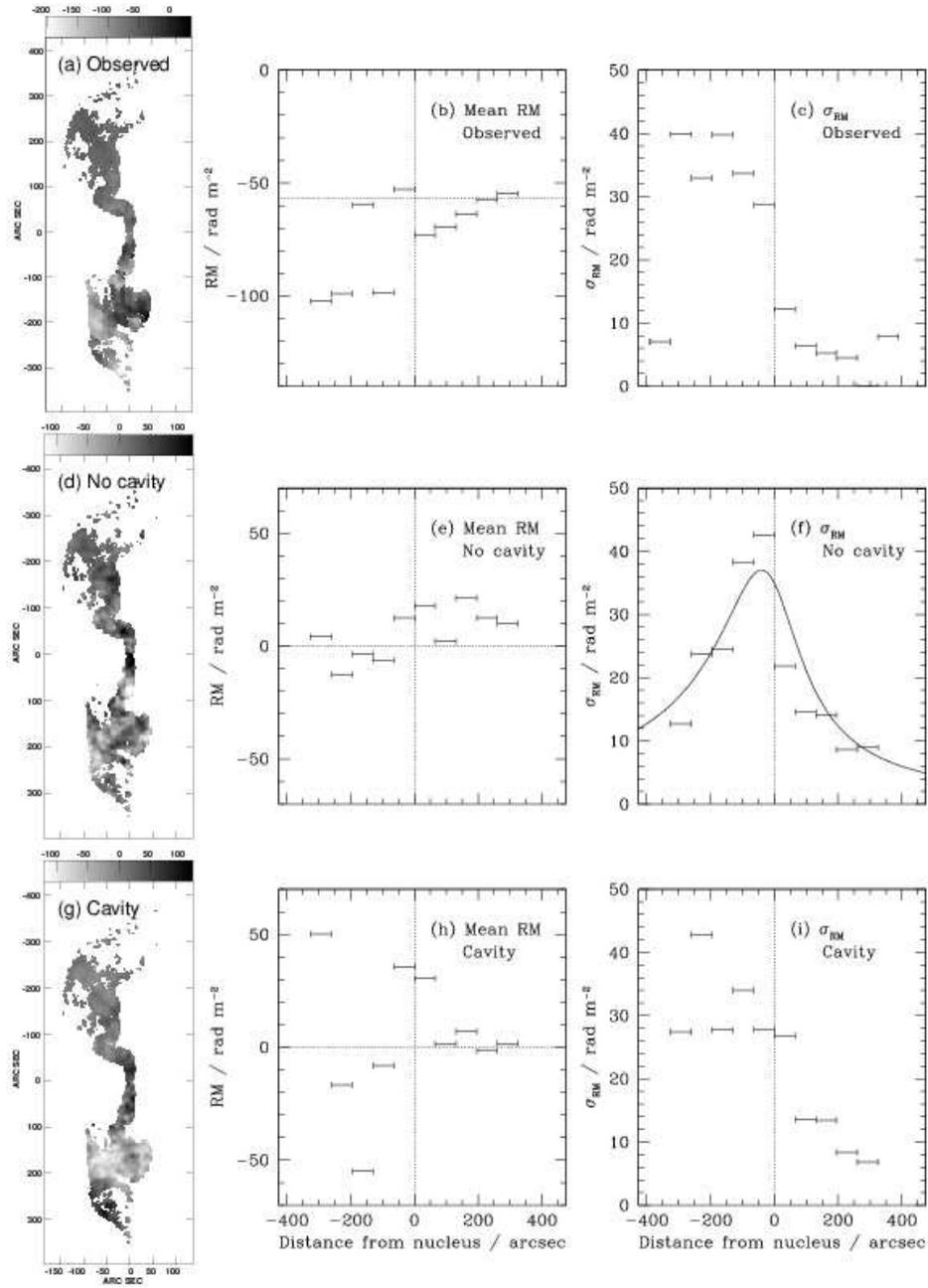}
\caption{Comparison of RM data at a resolution of 5.5\,arcsec FWHM with
  three-dimensional simulations. Panels (a) -- (c): data; (d) -- (f) simulation
  with spherically symmetrical gas distribution (central rms magnetic field
  strength $\langle B_0^2 \rangle^{1/2}$ =
  0.21 nT); (g) -- (i): simulation with cavity. (a), (d) and (g) show RM
  images. (b), (e) and (h) show mean values over boxes of length 65 arcsec along
  the jet axis, (f) and (i) the corresponding rms
  variations. The curve in panel (f) is the variation predicted by a
  single-scale model.\label{fig:3d}}
\end{figure}

We quantify the spatial variation of RM using the structure function $S({\bf r})
= \langle [{\rm RM}({\bf r} + {\bf r}^\prime) - {\rm RM}({\bf
r}^\prime)]^2\rangle$, where ${\bf r}$ and ${\bf r}^\prime$ are separation
vectors in the plane of the sky and $\langle\rangle$ denotes an average over
${\bf r}^\prime$. Our data are consistent with isotropy, so we plot $S$ as a
function of scalar separation $r$ for five regions of 3C\,31 in
Fig.~\ref{fig:sf}.  In order to interpret these results, we start from a model
RM power spectrum $\hat{C}(k)$, where $k$ is the wave-number. Its Hankel
transform is the RM autocorrelation function $C(r)$, and the structure function
is $S(r) = 2[C(r)-C(0)]$. A novel feature of our analysis is that we include the
effects of the observing beam explicitly (this can be done straightforwardly
provided that the rotation across the beam is small, as is the case for our
observations). Our data are consistent with a power spectrum which has the same
form everywhere, but varying normalization. We have found two acceptable
functional forms for the power spectrum: $\hat{C}(k) \propto k^{-2.35}$ with a
high-frequency cut-off at $k = 0.5$\,arcsec$^{-1}$ (a scale of 12\,arcsec or
4\,kpc) and a broken power-law form with $\hat{C}(k) \propto k^{-11/3}$ (as
expected for Kolmogorov turbulence) for $k > 0.13$\,arcsec$^{-1}$ and $\propto
k^{-1.5}$ at larger scales. The predicted structure functions are very similar
if the effects of the beam are included. This may explain why earlier studies
have come to different conclusions regarding the index of the power spectrum on
the basis of RM analysis alone \citep{VE03,VE05,Murgia,Govoni}.  The cut-off
power-law model with $\hat{C}(k) \propto k^{-2.35}$ predicts significantly less
residual depolarization, however, in better agreement with our data for 3C\,31.

\section{Three-dimensional simulations}

The easiest way to explain an asymmetry in RM fluctuation amplitude that is
correlated with jet sidedness is to postulate that the rotation arises in a
large-scale gas component surrounding the source -- most plausibly the
group-scale hot component. The asymmetry is then simply due to the differing
path lengths through this gas to the approaching (brighter) and receding jets
\citep{L88}.  In order to model the asymmetry, we simulate the Faraday rotation
from an isotropic, random magnetic field with a power spectrum corresponding to
that derived from the RM structure-function analysis embedded in ionized gas
with a smooth density distribution (cf.\ \citealt{Murgia}).  We first assume a
spherically-symmetric density distribution with a core radius of 150\,arcsec as
fit to {\sc rosat} data for the 3C\,31 group by \citet{KB99}.  An example RM
distribution for a radio source of negligible thickness inclined by 52$^\circ$
to the line of sight \citep{LB02} is shown in Figs~\ref{fig:3d}(d) -- (f).  The
rms central magnetic field strength for this model is $\langle B_0^2
\rangle^{1/2}$ = 0.21\,nT (2.1\,$\mu$G).  The predicted RM distribution shows an
asymmetry, but there is no sharp change at the nucleus and the profile falls too
rapidly with distance on the counter-jet side. The most plausible reason for the
discrepancy is that the radio lobes have displaced the surrounding gas, which is
therefore highly non-spherical. Cavities in the X-ray-emitting gas associated
with radio lobes are indeed observed in similar sources such as 3C\,449
\citep{Croston}. We assume that the contents of the cavity produce negligible
Faraday rotation. If we model the cavity as initially conical, with a
half-opening angle of $\approx$55$^\circ$ within 100\,arcsec of the nucleus and
thereafter cylindrical, we can approximately reproduce the RM distribution
(Figs~\ref{fig:3d}g -- i).  Such a cavity would not have been apparent in the
existing X-ray data \citep{KB99}, but should be detectable with XMM-Newton.

\acknowledgements %%% Text of acknowledgements runs on after this command.
The National Radio Astronomy Observatory is a
facility of the National Science Foundation operated under cooperative agreement
by Associated Universities, Inc.

%%% THE BIBLIOGRAPHY
%%%
%%% CONSULT SECTION 3 OF "INSTRUCTIONS FOR AUTHORS" FOR HOW TO USE NATBIB.
%%% AUTHORS ARE ENCOURAGED TO USE EITHER THE "THEBIBLIOGRAPY" ENVIRONMENT
%%% BY UNCOMMENTING (DELETING THE "%" SYMBOL) THE COMMANDS BELOW, OR BY
%%% USING THE BIBTEX ENVIRONMENT. TO FIND OUT WHICH IS APPLICABLE TO YOUR
%%% CONTRIBUTION, CONSULT THE VOLUME EDITORS FOR YOUR PROCEEDINGS.
%%%

\end{document}